\documentclass[twocolumn,amsmath,showpacs,amsfonts,aps,prc,floatfix]{revtex4}
\usepackage{graphicx}
\usepackage{bm}
\newcommand{\la}{\langle}
\newcommand{\ra}{\rangle}
\newcommand{\etal}{{\em et al.}}
\begin{document}

%%\draft
\title{Energy scan by $\phi$ mesons and threshold energy for the confinement-deconfinement phase transition}
 
\author{A. K. Chaudhuri}
\email[E-mail:]{akc@veccal.ernet.in}
\affiliation{Variable Energy Cyclotron Centre, 1/AF, Bidhan Nagar, 
Kolkata 700~064, India}

\begin{abstract}
We argue that the ratio of $\phi$ mesons multiplicity over cube of the mean $p_T$ is proportional to the degeneracy of the medium produced in ultra-relativistic heavy ion collisions. The ratio extracted from the existing $\phi$ meson data   in the energy range $\sqrt{s}$=6.3-200 GeV, indicate
that   beyond a threshold energy $\sqrt{s}_{th}=15.74\pm 8.10$ GeV, the medium crosses over from a confined phase to a deconfined phase. 
\end{abstract}

\pacs{25.75.-q, 25.75.Nq, 25.75.Ag} 

\date{\today}  

\maketitle

%%%%%%%%%%%%%%%%%%%%%%%%%%%%%%%%%%%%%%%%%%%%%%%%%%%%%%%%%%%%%%%%%%
%\section{Introduction} \label{sec1}
%%%%%%%%%%%%%%%%%%%%%%%%%%%%%%%%%%%%%%%%%%%

Lattice QCD predicts  \cite{lattice,Cheng:2007jq} that in ultra-relativistic heavy ion collisions,  a confinement-deconfinement phase transition can occur, producing    a new state of matter, Quark-Gluon Plasma (QGP). QGP is a collective state where color degrees of freedom become manifest over nuclear rather than hadronic volume.  Recent experiments  \cite{BRAHMSwhitepaper,PHOBOSwhitepaper,PHENIXwhitepaper,STARwhitepaper} have  produced convincing evidences for a confinement-deconfinement phase transition  in $\sqrt{s}$=200 GeV Au+Au collisions at RHIC.   
% Recent experiments at RHIC \cite{BRAHMSwhitepaper,PHOBOSwhitepaper,PHENIXwhitepaper,STARwhitepaper} have  produced convincing evidences that in $\sqrt{s}$=200 GeV Au+Au collisions, a confinement-deconfinement phase transition has occurred.   
%
One naturally wonders, whether or not there is a threshold energy for the transition. 
One of the aims of the STAR's energy scan programme  at RHIC is to determine the threshold energy for the confinement-deconfinement transition \cite{Caines:2009yu,Odyniec:2008zz}. STAR proposes to study nuclear collisions at $\sqrt{s}$=5, 7.7, 11.5, 17.3, 27 and 39 GeV. With the existing data at $\sqrt{s}$=62,130 and 200 GeV, STAR will scan a   large energy range, $\sqrt{s}$=5-200 GeV.

For long, strangeness enhancement is considered as a signature of QGP formation  \cite{Koch:1986ud}. In QGP environment, $gg\rightarrow s\bar{s}$ is abundant.
If not annihilated before hadronisation, early produced strange and anti-strange quarks will coalesce in to form strange hadrons and 
compared to elementary pp collisions, strange particle production will be enhanced.  
However, strangeness enhancement could also be obtained in a purely hadronic scenario, mainly due to the 'volume effect'  \cite{Rafelski:1980gk,Cleymans:1998yb,Hamieh:2000tk,Tounsi:2002nd}. Strangeness production in small volume elementary pp collisions can be 'canonically' suppressed due to 'strict' strangeness conservation \cite{Rafelski:1980gk,Cleymans:1998yb,Hamieh:2000tk,Tounsi:2002nd}.
In bigger volume AA collisions, locally, strangeness conservation condition can be relaxed to produce strange particles. In the language of statistical mechanics, while canonical ensemble is applicable in pp collisions, grand canonical ensemble is more appropriate in heavy ion collisions. Additionally, strange particle phase space appears to be under-saturated in elementary pp or peripheral heavy ion collisions \cite{Becattini:2003wp,Becattini:2005xt}.  

As noted in \cite{Mohanty:2009tz}, several unique features of $\phi$ mesons 
make it an ideal probe to investigate medium properties in heavy ion collisions.
They are hidden strange particle and are not affected by canonical suppression. Also they are not affected by resonance decay, have
  both hadronic and leptonic decay channels,   mass and width are not modified in a medium \cite{Alt:2008iv},  etc. Incidentally, experimental data for $\phi$ meson production in nuclear collisions over a large energy range $\sqrt{s}$=6-200 GeV exists. NA49 collaboration measured $\phi$ meson in 20A, 30A, 40A, 80A and 158A GeV Pb+Pb collisions \cite{Alt:2008iv}. The centre of mass energies are, 6.3, 7.6, 8.3, 12.3 and 17.3 GeV respectively. STAR collaboration measured $\phi$ mesons in Au+Au and Cu+Cu collisions at RHIC \cite{Abelev:2007rw,:2008fd,Abelev:2008zk}. In Au+Au collisions 
$\phi$ mesons are measured at three energies, $\sqrt{s}$=62, 130 and 200 GeV, and  in 
Cu+Cu collisions at $\sqrt{s}$=62 GeV and 200 GeV.  In the following we argue that the  existing $\phi$ meson data do indicate a threshold energy for the confinement-deconfinement phase transition.

 \begin{figure}[t]
%\vspace{0.3cm}
\center
 \resizebox{0.4\textwidth}{!}{%
  \includegraphics{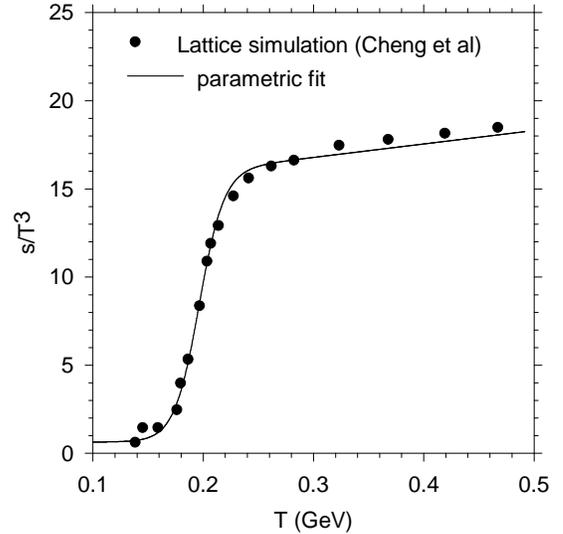}
}
\caption{ Black circles are lattice simulation \cite{Cheng:2007jq} for $s/T^3$, entropy density over cube of the temperature. The solid line is a parameterisation of lattice simulation of $s/T^3$.}  \label{F1}
\end{figure} 

Let us note the most distinguishing feature between a confined and a deconfined state. Effective degrees of freedom of a deconfined phase is considerably larger than that of a confined phase. As an example, in Fig.\ref{F1}, a recent lattice simulation \cite{Cheng:2007jq} for the temperature dependence of entropy density over cube of the temperature ($s/T^3$) is shown. The simulation was done with two light flavor quarks and a heavy strange quark, with almost physical quark masses. $s/T^3$ is approximately proportional to the degeneracy of the medium.   
From high temperature phase to low temperature phase, over  a narrow temperature range, degeneracy of the medium  drops rapidly, by factor of 4-5.

 \begin{figure}[t]
%\vspace{0.3cm}
\center
 \resizebox{0.4\textwidth}{!}{%
  \includegraphics{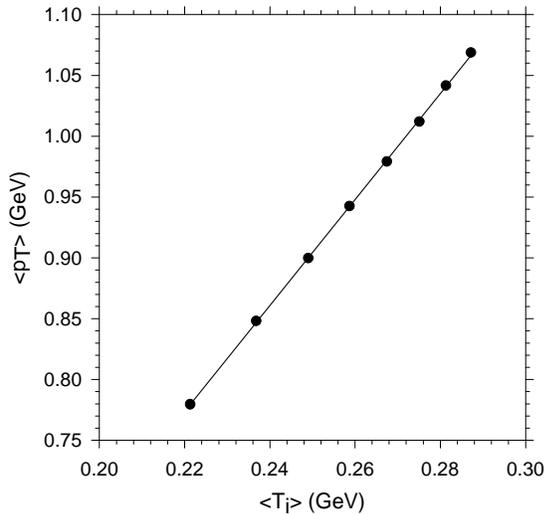}
}
\caption{The black circles hydrodynamic model simulations for $\phi$ mesons mean $p_T$ 
as a function of spatially averaged initial temperature $<T_i>$. The solid line indicate that
in a hydrodynamic model,  $\phi$ mesons mean $p_T$ depend linearly on the average initial temperature.}  \label{F2}
\end{figure} 

Can we construct an experimental observable equivalent to $s/T^3$, i.e. degeneracy of the medium? Variation of the observable with collision energy then can answer, whether or not  threshold energy exists for the confinement-deconfinement transition. We argue that the experimental observable, $\frac{\la N^\phi\ra}{\la p^\phi_T\ra^3}$, ratio of the $\phi$ meson multiplicity ($\la N^\phi\ra$) and cube of the $\phi$ mesons mean transverse momentum ($\la p^\phi_T\ra$) is effectively proportional to $s/T^3$, i.e.  the degeneracy   of the medium. The argument is based on three assumptions: (i) viscous effects are small in relativistic  energy heavy ion collisions, (ii) $\phi$ mesons multiplicity is proportional to initial entropy density and (iii) $\phi$ mesons mean $p_T$ is proportional to initial temperature. Let us examine the assumptions in detail. 
  At RHIC energy collisions,
experimental data at low $p_T< 1.5 GeV$ are consistent with ideal hydrodynamics \cite{QGP3}.
Assumption (i) is approximately valid in RHIC energy. However, viscous effects can be substantial at lower SPS energy  \cite{Kolb:2000fha,Ivanov:2009kn}.  Recently Petersen and Bleicher   \cite{Petersen:2009vx} studied elliptic flow at SPS energy. It was shown that initial conditions can have substantial effect on development of the elliptic flow. With proper initial conditions and with gradual freeze-out, elliptic flow data at SPS energy  can be explained in an ideal hydrodynamic model. In  \cite{Roy:2009bm}, $\phi$ mesons transverse momentum spectra, in the energy range $\sqrt{s}$=6-200 GeV, were analysed. Ideal hydrodynamics reasonably well explained the spectra. 
Assumption of small viscous effect in relativistic heavy ion collisions seems to be reasonable at SPS energy also. If viscous effects are small, then assumption (ii) i.e. $\phi$ meson multiplicity is proportional to initial entropy density, is also reasonable \cite{Hwa:1985xg}. It is generally believed that total multiplicity is proportional to final state entropy density. If viscous effects are small, entropy is not generated and initial and final state  entropy is same.
Experimental data on production cross-section of various particles and anti-particles show surprisingly good agreement with thermal abundances of a hadronic resonance gas \cite{BraunMunzinger:2001ip,Cleymans:1999st,Cleymans:1998fq}. The assumption is valid in a thermal model. However, some of the thermal models use 'strangeness under saturation ($\gamma_S$)' factor to explain the data \cite{Becattini:2003wp,Becattini:2005xt}.  $\gamma_S\approx$0.6 in low AGS energy. It increases with energy and at RHIC energy $\gamma_S\approx$ 1. The assumption may not hold in
thermal models with strangeness under saturation factor.
 The assumption (iii), $\phi$ mesons mean $p_T$ is proportional to the initial temperature is approximately valid in an ideal hydrodynamic model. In Fig.\ref{F2}, (ideal) hydrodynamic model predictions for $\phi$ mesons mean $p_T$ in a central (0-10\%) Au+Au collisions, are shown as a function of    'spatially' averaged initial temperature.
 At the initial time $\tau_i$=0.6 fm, initial energy density is assumed to be distributed as \cite{QGP3}

 \begin{figure}[t]
%\vspace{0.9cm}
\center
 \resizebox{0.4\textwidth}{!}{%
  \includegraphics{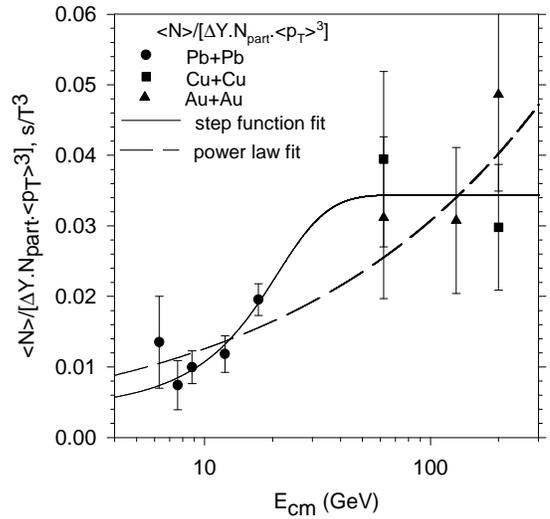}
}
\caption{   The filled symbols are the ratio of $\phi$ meson multiplicity over the cube of the mean $p_T$, normalised by participant number ($N_{part}$) and rapidity gap ($\Delta Y$), in Pb+Pb, Cu+Cu and Au+Au collisions, as a function of the collision energy. The solid line is a fit to the ratio by the analytical form for the step function Eq.\ref{eq2}, the dotted line is a power law fit.  }  \label{F3}
\end{figure}  

\begin{equation} \label{eq1}
\varepsilon({\bf b},x,y)=\varepsilon_i[0.75 N_{part}({\bf b},x,y) +0.25 N_{coll}({\bf b},x,y)].
\end{equation}

In Eq.\ref{eq1}, {\bf b} is the impact parameter of the collision and $N_{part}$ and $N_{coll}$ are the transverse profile of the average number of  participants and average number of binary collisions respectively, calculated in a Glauber model. Initial fluid velocity  is zero, $v_x(x,y)=v_y(x,y)=0$. Hydrodynamic equations are solved
with the code AZHYDRO-KOLKATA, detail of which can be found in \cite{Chaudhuri:2008sj,Chaudhuri:2008ed,Chaudhuri:2009uk}.
We have used an equation of state, with a cross-over transition from QGP phase to hadronic phase at temperature $T_{co}$=196 MeV \cite{Chaudhuri:2009uk}. For a set of central energy density, $\varepsilon_i$, $\phi$ mesons mean $p_T$'s are calculated at the freeze-out temperature $T_F$=150 MeV. In ideal hydrodynamics, a linear relation between   $\la p^\phi_T\ra$ and average initial temperature $\la T_i\ra$ is accurately observed. 
It is interesting to note that the relation $\la p^\phi_T\ra \propto \la T_i \ra$ will not be valid, say for the pions or   for the strange mesons $K^+/K^-$. Pions and kaons are largely affected by resonance decay, which spoils the relation. Resonances do not contribute to $\phi$ meson production making the relation works.  
  
 \begin{figure}[t]
%\vspace{0.9cm}
\center
 \resizebox{0.4\textwidth}{!}{%
  \includegraphics{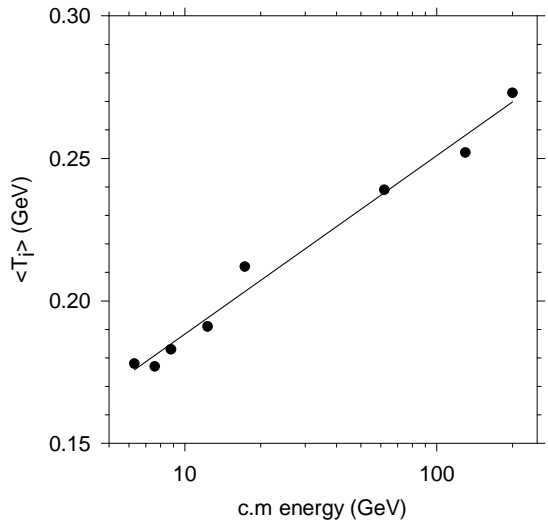}
}
\caption{The filled circles are spatially averaged initial temperature of the fluid, obtained from an ideal hydrodynamic model analysis of $\phi$ meson $p_T$ spectra in the energy range $\sqrt{s}$=6-200 GeV. The solid line is a fit to the average temperature.   }  \label{F4}
\end{figure}   
  
NA49 and STAR collaboration have tabulated $\phi$ multiplicity and mean $p_T$ in Pb+Pb and Au+Au/Cu+Cu collisions 
in the energy range $\sqrt{s}$= 6.3-200 GeV \cite{Alt:2008iv,Abelev:2007rw,:2008fd,Abelev:2008zk}.  
  In Fig.\ref{F3}, the experimental ratio $\la N^\phi\ra/\la p^\phi_T\ra^3$ is shown as a function of the collision energy.  In Au+Au/Cu+Cu collisions, $\phi$ mesons are measured in a number of collisions centrality. Presently, we chose the most central ones.
To account for the differences in system size, collision centrality and rapidity gap ($\Delta Y$) in NA49 and STAR data sets, $\phi$ multiplicity is normalised by the factor $0.5N_{part} \Delta Y$.  
Uncertainty in the experimental ratio $\la N^\phi\ra/\la p^\phi_T\ra^3$ is large due to the large experimental error in determination of $\phi$ meson multiplicity and mean $p_T$. For example, in low SPS energy, uncertainty in $\phi$ multiplicity is as large as $\sim$20-40\%. At RHIC energy also, $\la N^\phi \ra$ is determined only within $\sim$ 10-20\% accuracy. 
Mean $p_T$ is determined more accurately, in the energy range $\sqrt{s}$=6.3-200 GeV, uncertainty in $<p^\phi_T>$ is less than 10\%. Though error bars are large, the trend of the ratio, $R=\la N^\phi\ra/[.5N_{part}\Delta Y \la p^\phi_T\ra^3]$ as a function of the collision energy is evident. Mimicking the temperature dependence of $s/T^3$, the ratio sharply rises from low SPS energy to RHIC energy.  In Fig.\ref{F3}, the solid line is a fit to the ratio with an analytical form for the step function,

\begin{equation} \label{eq2}
R=\alpha[1+tanh\frac{\sqrt{s}-\sqrt{s}_{th}}{\Delta\sqrt{s}_{th}}].
\end{equation}

Analytical form Eq.\ref{eq2}, well explain the data, $\chi^2/N=0.3$.
Fitted values are $\alpha=0.017\pm 0.009$, $\sqrt{s}_{th}=15.74\pm 8.10$GeV and
$\Delta\sqrt{s}_{th}=14.52\pm 14.93$ GeV. Threshold energy can be determined only within $\sim$50\% accuracy, the width of the transition is uncertain by $\sim$100\%. 
One  note that presently, no data exist in the energy range 17.3-62 GeV, between the top of SPS energy and bottom of the RHIC energy.     STAR energy scan programme will fill up the gap. Threshold energy and width of the   transition can be determined more accurately.

 \begin{figure}[t]
%\vspace{0.9cm}
\center
 \resizebox{0.4\textwidth}{!}{%
  \includegraphics{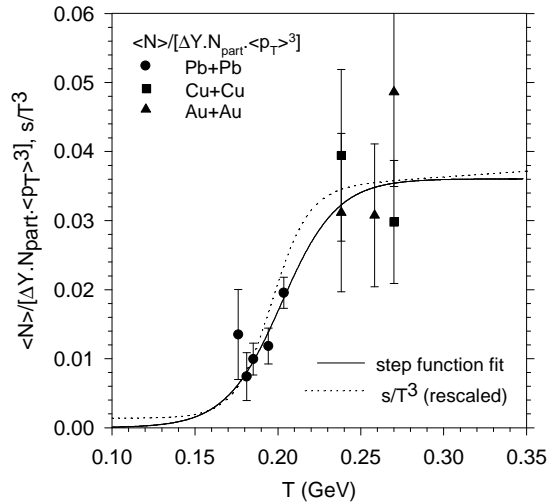}
}
\caption{The experimental ratio $R$ is plotted as a function of temperature. The solid line is a step function fit to the ratio. For comparison, we have also shown (rescaled) $s/T^3$  in lattice simulation by cheng \etal
 \cite{Cheng:2007jq}, the dotted line.}    \label{F5}
\end{figure}  

One may argue that fitting the ratio  $\la N^\phi\ra/\la p^\phi_T\ra^3$ by an analytical form for the step function is rather arbitrary, the ratio could as well be fitted by another form, without   any threshold energy. 
As an example, in Fig.\ref{F3}, we have shown a fit (the dashed line) to the ratio $R$, by a power law, $R=A \sqrt{s}^B$. Power law also explains the data, but with increased $\chi^2/N=0.92$.  Since  the step function fit as well as  the power law fit give $\chi^2/N <1$, from the $\chi^2$ analysis point of view, both the fits are equivalent and it can not be claimed that the experimental ratio  $\la N^\phi\ra/\la p^\phi_T\ra^3$ exhibit step function like behavior.
Indeed, one does wonder, whether it can be verified that the ratio  $\la N^\phi\ra/\la p^\phi_T\ra^3$ do indeed corresponds to $s/T^3$, i.e. the degrees of freedom of the medium, as argued here. 
Note that   $s/T^3$ is a function of temperature, not of collision energy. Can we convert the collision energy dependence of $\la N^\phi\ra/\la p^\phi_T\ra^3$ to temperature dependence? As noted earlier, in \cite{Roy:2009bm}, 
 $\phi$ meson $p_T$ spectra in the energy range 6-200 GeV were  analysed in an ideal  hydrodynamic model. Ideal hydrodynamics reasonably well explain the $p_T$ spectra of $\phi$ mesons. In Fig.\ref{F4}, spatially averaged initial temperature of the fluid as obtained in the hydrodynamic analysis \cite{Roy:2009bm}, are shown as 
a function of the collision energy. Spatially averaged initial temperature depend logarithmically on the collision energy,

 \begin{figure}[t]
 %\vspace{0.9cm}
\center
 \resizebox{0.4\textwidth}{!}{%
  \includegraphics{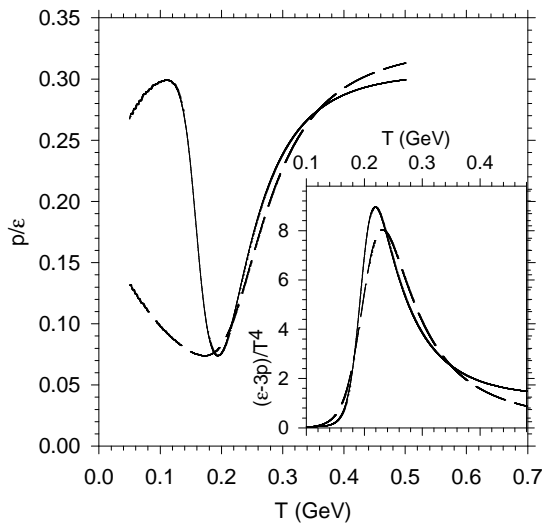}
}
\caption{The solid line is the square speed of sound ($c_s^2$) in lattice simulation  \cite{Cheng:2007jq}. The dashed line is $c_s^2$ in the model assuming the step function fit   to the experimental ratio $\la N^\phi\ra/\la p^\phi_T\ra^3$
 is proportional to $s/T^3$, the entropy density over cube of the temperature. In the inset, trace anomaly $\frac{\varepsilon-3p}{T^4}$ in lattice simulation and in the model are plotted. Model trace anomaly is normalised by the factor $K=460$.}
   \label{F6}
\end{figure}  

\begin{equation} \label{eq3}
<T_i>=A+B \log \sqrt{s}
\end{equation}

\noindent with $A=0.126$, $B=0.063$. We use Eq.\ref{eq3} to convert the collisional energy dependence of the ratio, $\la N^\phi\ra/\la p^\phi_T\ra^3$ in to temperature dependence. Fig.\ref{F5} shows  the ratio  as a function of  temperature. As a function of temperature also, the ratio show a jump from low temperature to the high temperature. The ratio is still fitted with a step function, with $T_{th}=202 \pm 23 $MeV, $\Delta T_{th}=33 \pm 37 $ MeV. The  threshold temperature, $T_{th}$=202 MeV,   is very close to the cross-over temperature $T_{co}$=197$\pm$ 3 MeV, obtained in the recent lattice simulation  \cite{Cheng:2007jq}.
The dotted line in Fig.\ref{F5}, is a parametric representation for the entropy density in lattice simulation  \cite{Cheng:2007jq}. It has been scaled by a factor $K=460$.  
Experimental ratio $\la N^\phi\ra/\la p^\phi_T\ra^3$ closely agree with lattice simulated $s/T^3$ and also with the step function fit to the ratio. Only in the temperature range 210-240 MeV, where data are sparse, rescaled $s/T^3$  overestimate the step function fit by 10-20\%. Close similarity between lattice simulated $s/T^3$ and experimental ratio $\la N^\phi\ra/\la p^\phi_T\ra^3$ do support our argument that the ratio $\la N^\phi\ra/\la p^\phi_T\ra^3$ is proportional to $s/T^3$ or the degeneracy of the medium. To confirm that the ratio $\la N^\phi\ra/\la p^\phi_T\ra^3$ is indeed proportional to $s/T^3$, 
using the usual thermodynamic relations,

\begin{eqnarray}
p(T)&=&\int_0^T s(T^\prime)dT^\prime \\
\varepsilon(T)&=&Ts-p,
\end{eqnarray}

\noindent
we obtained pressure and energy density for the step function fit to the ratio $\la N^\phi\ra/\la p^\phi_T\ra^3$  and  computed square speed of sound
($c_s^2=\frac{dp}{d\varepsilon}$). In Fig.\ref{F6},
  the dashed line is $c_s^2$, obtained from the step function fit to the experimental ratio $\la N^\phi\ra/\la p^\phi_T\ra^3$. The solid line is $c_s^2$ from a parametric representation of the lattice simulations for $s/T^3$. For $T>$200 MeV, $c_s^2$ from the step function fit to the experimental ratio closely agree with $c_s^2$ in lattice simulation. $c_s^2$ in lattice simulation show a dip at $T\approx$200 MeV.
$c_s^2$ from the step function fit to the ratio show a dip at $T\approx$180 MeV.
However, at low temperature, $c^2_s$ is comparatively larger in lattice simulation
than in the step function fit to the ratio. We note that   experimental data for the ratio  $\la N^\phi\ra/\la p^\phi_T\ra^3$ do not exit below temperature $T\approx$170 MeV and the step function fit may not be accurate at low temperature.  
In the inset of Fig.\ref{F6}, the trace anomaly $\frac{\varepsilon-3p}{T^4}$,
is shown as a function of the temperature. The solid line is the trace anomaly in lattice simulation, the dashed  line is that from the step function fit to $\la N^\phi\ra/\la p^\phi_T\ra^3$. Trace anomaly from the step function is normalised by a factor $K=460$. Normalised trace anomaly closely agree with the lattice simulated value. We may note here that if the ratio $\la N^\phi\ra/\la p^\phi_T\ra^3$
 is fitted with a power law,  $c_s^2$ and trace anomaly from the   fit do not resemble lattice simulation results. For example, square speed of sound does not show a dip or 
the trace anomaly does not show a peak.

To conclude, we have argued that the ratio of $\phi$ mesons multiplicity over cube of the mean $p_T$ corresponds to the degeneracy of the medium produced in relativistic heavy ion collisions and can signal the confinement-deconfinement phase transition. The ratio constructed from the existing data \cite{Alt:2008iv,Abelev:2007rw,:2008fd,Abelev:2008zk} in the energy range $\sqrt{s}$=6.3-200 GeV, even though  have large error bars, do show rapid rise from low SPS energy to high RHIC energy.
Using a result of hydrodynamical analysis of $\phi$ mesons $p_T$ spectra \cite{Roy:2009bm}, that the average initial temperature depend logarithmically on the collision energy,
 the collision energy dependence of the ratio is converted in to temperature dependence.
The ratio as a function of the temperature closely corresponds to lattice simulations for $s/T^3$, the entropy density over cube of the temperature. From a step function fit to the ratio we also have extracted the threshold energy,  $\sqrt{s}_{th}=15.74\pm 8.10$ GeV, for the confinement-deconfinement phase transition.


\begin{thebibliography}{99}
\bibitem{lattice} 
Karsch F, Laermann E, Petreczky P, Stickan S and Wetzorke I, 
2001 {\it Proccedings of NIC Symposium} (Ed. H. Rollnik and D. Wolf, John 
von Neumann Institute for Computing, J\"ulich, NIC Series, vol.9, 
ISBN 3-00-009055-X, pp.173-82,2002.)

%\cite{Cheng:2007jq}
\bibitem{Cheng:2007jq}
  M.~Cheng {\it et al.},
  %``The QCD Equation of State with almost Physical Quark Masses,''
  Phys.\ Rev.\  D {\bf 77}, 014511 (2008)
  [arXiv:0710.0354 [hep-lat]].
  %%CITATION = PHRVA,D77,014511;%%


\bibitem{BRAHMSwhitepaper}
 BRAHMS Collaboration, I. Arsene {\it et al.},  
%"Quark-gluon plasma and the color glass condensate at RHIC?  
% The prespective from the BRAHMS experiment" 
Nucl. Phys. A {\bf 757}, 1 (2005). 
%nucl-ex/0410020 
 
\bibitem{PHOBOSwhitepaper} 
PHOBOS Collaboration,  B. B. Back {\it et al.},  
%"The PHOBOS Perspective on Discoveries at RHIC'' 
Nucl. Phys. A {\bf 757}, 28 (2005). 
%nucl-ex/0410020 
 
\bibitem{PHENIXwhitepaper} 
PHENIX Collaboration, K.~Adcox {\it et al.}, 
%``Formation of dense partonic matter in relativistic nucleus nucleus 
%collisions at RHIC: Experimental evaluation by the PHENIX  collaboration,'' 
Nucl. Phys. A {\bf 757} 184 (2005).  
%arXiv:nucl-ex/0410003. 
%%CITATION = NUCL-EX 0410003;%% 
  
\bibitem{STARwhitepaper} 
STAR Collaboration, J. Adams {\it et al.}, 
%''Experimental and theoretical challenges in the search for the quark  
%  gluon plasma: The STAR Collaboration's critical assessment of the  
%  evidence from RHIC collisions'' 
Nucl. Phys. A {\bf 757} 102 (2005).% [arXiv:nucl-ex/0501009]. 
%%CITATION = NUCL-EX 0501009;%% 

%\cite{Caines:2009yu}
\bibitem{Caines:2009yu}
  H.~Caines  [STAR Collaboration],
  %``The RHIC Beam Energy Scan - STAR'S Perspective,''
  arXiv:0906.0305 [nucl-ex].
  %%CITATION = ARXIV:0906.0305;%%


%\cite{Odyniec:2008zz}
\bibitem{Odyniec:2008zz}
  G.~Odyniec  [STAR Collaboration],
  %``STAR physics program and technical challenges with the RHIC energy scan
  %with Au + Au collisions,''
  J.\ Phys.\ G {\bf 35}, 104164 (2008).
  %%CITATION = JPHGB,G35,104164;%%
 
 %\cite{Koch:1986ud}
\bibitem{Koch:1986ud}
  P.~Koch, B.~Muller and J.~Rafelski,
  %``Strangeness In Relativistic Heavy Ion Collisions,''
  Phys.\ Rept.\  {\bf 142}, 167 (1986).
  %%CITATION = PRPLC,142,167;%%    

%\cite{Rafelski:1980gk}
\bibitem{Rafelski:1980gk}
  J.~Rafelski and M.~Danos,
  %``The Importance Of The Reaction Volume In Hadronic Collisions,''
  Phys.\ Lett.\  B {\bf 97}, 279 (1980).
  %%CITATION = PHLTA,B97,279;%%  
  
%\cite{Cleymans:1998yb}
\bibitem{Cleymans:1998yb}
  J.~Cleymans, H.~Oeschler and K.~Redlich,
  %``Influence Of Impact Parameter On Thermal Description Of Relativistic Heavy
  %Ion Collisions At (1-2) A-Gev,''
  Phys.\ Rev.\  C {\bf 59}, 1663 (1999)
  [arXiv:nucl-th/9809027].
  %%CITATION = PHRVA,C59,1663;%%  

 %\cite{Hamieh:2000tk}
\bibitem{Hamieh:2000tk}
  S.~Hamieh, K.~Redlich and A.~Tounsi,
  %``Canonical description of strangeness enhancement from p A to Pb Pb
  %collisions,''
  Phys.\ Lett.\  B {\bf 486}, 61 (2000)
  [arXiv:hep-ph/0006024].
  %%CITATION = PHLTA,B486,61;%%
%\cite{Tounsi:2002nd}
\bibitem{Tounsi:2002nd}
  A.~Tounsi, A.~Mischke and K.~Redlich,
  %``Canonical aspects of strangeness enhancement,''
  Nucl.\ Phys.\  A {\bf 715}, 565 (2003)
  [arXiv:hep-ph/0209284].
  %%CITATION = NUPHA,A715,565;%%  
 
%\cite{Becattini:2003wp}
\bibitem{Becattini:2003wp}
  F.~Becattini, M.~Gazdzicki, A.~Keranen, J.~Manninen and R.~Stock,
  %``Study of chemical equilibrium in nucleus nucleus collisions at AGS and  SPS
  %energies,''
  Phys.\ Rev.\  C {\bf 69}, 024905 (2004)
  [arXiv:hep-ph/0310049].
  %%CITATION = PHRVA,C69,024905;%%
 %\cite{Becattini:2005xt}
\bibitem{Becattini:2005xt}
  F.~Becattini, J.~Manninen and M.~Gazdzicki,
  %``Energy and system size dependence of chemical freeze-out in  relativistic
  %nuclear collisions,''
  Phys.\ Rev.\  C {\bf 73}, 044905 (2006)
  [arXiv:hep-ph/0511092].
  %%CITATION = PHRVA,C73,044905;%%    
  
  
  
 %\cite{Mohanty:2009tz}
\bibitem{Mohanty:2009tz}
  B.~Mohanty and N.~Xu,
  %``Probe the QCD phase diagram with \phi-mesons in high energy nuclear
  %collisions,''
  arXiv:0901.0313 [nucl-ex].
  %%CITATION = ARXIV:0901.0313;%%

 %\cite{Alt:2008iv}
\bibitem{Alt:2008iv}
  C.~Alt {\it et al.}  [NA49 collaboration],
  %``Energy dependence of phi meson production in central Pb+Pb collisions at
  %$\sqrt{s}_{NN}$ = 6 to 17 GeV,''
  Phys.\ Rev.\  C {\bf 78}, 044907 (2008)
  [arXiv:0806.1937 [nucl-ex]].
  %%CITATION = PHRVA,C78,044907;%%     
   
   %\cite{Abelev:2007rw}
\bibitem{Abelev:2007rw}
  B.~I.~Abelev {\it et al.}  [STAR Collaboration],
  %``Partonic flow and Phi-meson production in Au + Au collisions at
  %s(NN)**(1/2) = 200-GeV,''
  Phys.\ Rev.\ Lett.\  {\bf 99}, 112301 (2007)
  [arXiv:nucl-ex/0703033].
  %%CITATION = PRLTA,99,112301;%%
\bibitem{:2008fd}
  B.~I.~Abelev {\it et al.}  [STAR Collaboration],
  %``Measurements of $\phi$ meson production in relativistic heavy-ion
  %collisions at RHIC,''
  arXiv:0809.4737 [nucl-ex].
  %%CITATION = ARXIV:0809.$;%%
 %\cite{Abelev:2008zk}
\bibitem{Abelev:2008zk}
  B.~I.~Abelev {\it et al.}  [STAR Collaboration],
  %``Energy and system size dependence of \phi meson production in Cu+Cu and
  %Au+Au collisions,''
  Phys.\ Lett.\  B {\bf 673}, 183 (2009)
  [arXiv:0810.4979 [nucl-ex]].
  %%CITATION = PHLTA,B673,183;%% 

  
 \bibitem{QGP3}
P.~F. Kolb and U. Heinz,
in {\it Quark-Gluon Plasma 3}, edited by R.~C. Hwa and 
X.-N. Wang (World Scientific, Singapore, 2004), p.~634.
%%CITATION = NUCL-TH 0305084;%%

%\cite{Kolb:2000fha}
\bibitem{Kolb:2000fha}
  P.~F.~Kolb, P.~Huovinen, U.~W.~Heinz and H.~Heiselberg,
  %``Elliptic flow at SPS and RHIC: From kinetic transport to  hydrodynamics,''
  Phys.\ Lett.\  B {\bf 500}, 232 (2001)
  [arXiv:hep-ph/0012137].
  %%CITATION = PHLTA,B500,232;%%
\cite{Ivanov:2009kn}
\bibitem{Ivanov:2009kn}
  Yu.~B.~Ivanov, I.~N.~Mishustin, V.~N.~Russkikh and L.~M.~Satarov,
  %``Elliptic Flow and Dissipation at AGS--SPS Energies,''
  Phys.\ Rev.\  C {\bf 80}, 064904 (2009)
  [arXiv:0907.4140 [nucl-th]].
  %%CITATION = PHRVA,C80,064904;%%
  %\cite{Petersen:2009vx}
\bibitem{Petersen:2009vx}
  H.~Petersen and M.~Bleicher,
  %``Ideal hydrodynamics and elliptic flow at SPS energies: Importance of the
  %initial conditions,''
  Phys.\ Rev.\  C {\bf 79}, 054904 (2009)
  [arXiv:0901.3821 [nucl-th]].
  %%CITATION = PHRVA,C79,054904;%%

 %\cite{Roy:2009bm}
\bibitem{Roy:2009bm}
  V.~Roy and A.~K.~Chaudhuri,
  %``A lattice based equation of state and $\phi$ meson production in
  %$\sqrt{s}$=6-200 GeV Pb+Pb and Au+Au collisions,''
  arXiv:0911.4556 [nucl-th].
  %%CITATION = ARXIV:0911.4556;%%
  
%\cite{Hwa:1985xg}
\bibitem{Hwa:1985xg}
  R.~C.~Hwa and K.~Kajantie,
  %``Diagnosing Quark Matter By Measuring The Total Entropy And The Photon Or
  %Dilepton Emission Rates,''
  Phys.\ Rev.\  D {\bf 32}, 1109 (1985).
  %%CITATION = PHRVA,D32,1109;%%





 


%\cite{BraunMunzinger:2001ip}
\bibitem{BraunMunzinger:2001ip}
  P.~Braun-Munzinger, D.~Magestro, K.~Redlich and J.~Stachel,
  %``Hadron production in Au Au collisions at RHIC,''
  Phys.\ Lett.\  B {\bf 518}, 41 (2001)
  [arXiv:hep-ph/0105229].
  %%CITATION = PHLTA,B518,41;%%
  
  %\cite{Cleymans:1999st}
\bibitem{Cleymans:1999st}
  J.~Cleymans and K.~Redlich,
  %``Chemical and thermal freeze-out parameters from 1-A-GeV to 200-A-GeV,''
  Phys.\ Rev.\  C {\bf 60}, 054908 (1999)
  [arXiv:nucl-th/9903063].
  %%CITATION = PHRVA,C60,054908;%%
  
%\cite{Cleymans:1998fq}
\bibitem{Cleymans:1998fq}
  J.~Cleymans and K.~Redlich,
  %``Unified description of freeze-out parameters in relativistic heavy ion
  %collisions,''
  Phys.\ Rev.\ Lett.\  {\bf 81}, 5284 (1998)
  [arXiv:nucl-th/9808030].
  %%CITATION = PRLTA,81,5284;%%
  
%\cite{Chaudhuri:2008sj}
\bibitem{Chaudhuri:2008sj} A.~K.~Chaudhuri,
  %``Viscous fluid dynamics in Au+Au collisions at RHIC,''
 arXiv:0801.3180 [nucl-th].
  %%CITATION = ARXIV:0801.3180;%%  
  
  %\cite{Chaudhuri:2008ed}
\bibitem{Chaudhuri:2008ed}
  A.~K.~Chaudhuri,
  %``Dissipative hydrodynamics and heavy ion collisions,''
  J.\ Phys.\ G {\bf 35}, 104015 (2008)
  [arXiv:0804.3458 [hep-th]].
  %%CITATION = JPHGB,G35,104015;%%


    
  
%\cite{Chaudhuri:2009uk}
\bibitem{Chaudhuri:2009uk}
  A.~K.~Chaudhuri,
  %``Nearly perfect fluid in Au+Au collisions at RHIC,''
  Phys.\ Lett.\  B {\bf 681}, 418 (2009)
  [arXiv:0909.0391 [nucl-th]].
  %%CITATION = PHLTA,B681,418;%%
  
\end{thebibliography}
\end{document}